\documentclass[a4paper,12pt]{article} 
\usepackage{amssymb,amsthm,latexsym}
\newtheorem{theorem}{Theorem} 
 
\title{EUCLIDEAN FIELD THEORY} 

\author{
Francesco Guerra\footnote{\ e-mail: {\tt francesco.guerra@roma1.infn.it}} \\ 
{\small{\itshape Dipartimento di Fisica, Universit\`a di Roma ``La Sapienza''}} \\ 
{\small {\itshape INFN, Sezione di Roma1, Piazzale A. Moro 2, 00185 Roma, Italy}}
}

\date{\today} 

\begin{document}

\maketitle 

\section{Introduction} 

In this review, we consider Euclidean field theory as a formulation of quantum field theory 
which lives in some Euclidean space, and is expressed in probabilistic terms. Methods arising 
from Euclidean field theory have been introduced in a very successful way in the study of the 
concrete models of Constructive Quantum Field Theory.

Euclidean field theory was initiated  by Schwinger
\cite{Schwinger} and Nakano \cite{Nakano}, who proposed to study the vacuum
expectation values of field products analytically continued into the
Euclidean region (Schwinger functions), where the first three (spatial)
coordinates of a world point are real and the last one (time) is purely
imaginary (Schwinger points). The possibility of introducing Schwinger
functions, and their invariance under the Euclidean group are immediate
consequences of the by now classic formulation of quantum field theory in
terms of vacuum expectation values given by Wightman \cite{SW}. The
convenience of dealing with the Euclidean group, with its positive definite
scalar product, instead of the Lorentz group is evident, and has been
exploited by several authors, in different contexts.

The next step was made by Symanzik \cite{Sym}, who realized that Schwinger
functions for Boson fields have a remarkable positivity property, allowing
to introduce Euclidean fields on their own sake. Symanzik also pointed out
an analogy between Euclidean field theory and classical statistical
mechanics, at least for some interactions \cite{Sym1}.

This analogy was
successfully extended, with a different interpretation, to all Boson
interaction by Guerra, Rosen and Simon \cite{GRS}, with the purpose of
using rigorous results of modern statistical mechanics for the study of
constructive quantum field theory, inside the program advocated by Wightman
\cite{Wigh}, and further pursued by Glimm and Jaffe (see \cite{GJ} for an
overall presentation).

The most dramatic advance of Euclidean theory was responsability of Nelson
\cite{Nel} \cite{Nel1}. He was able to isolate a crucial property of Euclidean fields (the
Markov property) and gave a set of conditions for Euclidean fields, which
allow to derive all properties of relativistic quantum fields satisfying
Wightman axioms. Nelson theory is very deep and rich of new ideas. After so
many years from the basic papers, we still lack a complete understanding of
the radical departure from the conventional theory afforded by Nelson
ideas, especially about their possible further developments.

By using Nelson scheme, in particular a very peculiar symmetry property, it
was very easy to prove \cite{G} the convergence of the ground state energy
density, and the Van Hove phenomenon in the infinite volume limit for two
dimensional Boson theories. A subsequent analysis \cite{GRS1} gave other
properties of the infinite volume limit of the theory, and allowed a
remarkable simplification in the proof of a very important regularity
property for fields, previously established by Glimm and Jaffe.

Since then all work on constructive quantum field theory has exploited in
different ways ideas coming from Euclidean field theory. Moreover, a very
important reconstruction theorem has been established by Osterwalder and
Schrader \cite{OS}, allowing reconstruction of relativistic quantum fields
from the Euclidean Schwinger functions, and avoiding the previously
mentioned Nelson reconstruction theorem, which is technically more
difficult to handle.

This paper is intended to be an introduction to the general structure of
Euclidean quantum field theory, and to some of the applications to
constructive quantum field theory. Our purpose is to show that, fifty years
after its introduction, the Euclidean theory is still interesting, both
from the point of view of technical application and physical
interpretation.

The paper is organized as follows. In Section 2, by considering simple
systems made of a single spinless relativistic particle, we introduce the
relevant structures in both Euclidean and Minkowski world. In particular, a
kind of (pre)Markov property is introduced already at the one particle level.

Next Section 3 contains the description of the procedure of second quantization on the one 
particle
structure. The free Markov field is introduced, and its crucial Markov property explained. By
following Nelson, we use probabilistic concepts and methods, whose relevance for constructive 
quantum
field theory became immediately more and more apparent. The very structure of classical 
statistical
mechanics for Euclidean fields is firmly based on these probabilistic methods. In
Section 4 we introduce the interaction and we show the connection between the Markov theory 
and the Hamiltonian theory, for
two-dimensional space-cutoff interacting scalar fields. In particular, we present the
Feynman-Kac-Nelson formula that gives an explicit expression of the semigroup generated by the
space-cutoff Hamiltonian in $\Phi o \kappa$ space. We deal also with some  applications to 
constructive quantum field theory.
Section 5 is dedicated to a short discussion about the physical interpretation of the theory. 
In particular we discuss the Osterwalder-Schrader reconstruction theorem on Euclidean 
Schwinger functions, and the Nelson reconstruction theorem on Euclidean fields. For the sake 
of completeness, we sketch the
main ideas of a proposal, advanced in \cite{GR}, according to which the Euclidean field theory 
can be
interpreted as a stochastic field theory in the physical Minkowski space-time. 

Our treatment will be as simple
as possible, by relying on the basic structural properties, and by describing methods of 
presumably
very long lasting power. The emphasis given to probabilistic methods, and to the statistical 
mechanics
analogy, is a result of the historical development. Our opinion is that not all possibility of
Euclidean field theory have been fully exploited yet, both from a technical and physical point 
of view.

\section {One particle systems}

A system made of only one relativistic scalar particle, of mass $m>0$, has a
quantum state space represented by the positive frequency solutions of the
Klein-Gordon equation. In momentum space, with points $p_\mu, \mu=0,1,2,3$, let us introduce 
the upper mass hyperboloid, characterized by the constraints $p^2\equiv p_0^2 - \sum_{i=1}^3 
p_i^2=m^2$, $p_0\ge m$, and the relativistic invariant measure on it, formally given by 
$d\mu(p)=\theta(p_0)\delta(p^2-m^2)dp$, where $\theta$ is the step function $\theta(x)=1$ if 
$x\ge 0$, and $\theta(x)=0$ otherwise, and $dp$ is the four-dimensional Lebesgue measure. The 
Hilbert space of quantum states $F$ is given by the square integrable functions on the mass 
hyperboloid equipped with the invariant measure $d\mu(p)$. Since in some reference frame the 
mass hyperboloid is uniquely characterized by the space values of the momentum ${\bf p}$, with 
the energy given by $p_0\equiv\omega({\bf p})=\sqrt{{\bf p}^2 + m^2}$, the Hilbert space $F$ 
of the states is in fact made
of those complex valued tempered distributions $f$ in the configuration space $R^3$, whose 
Fourier
transforms $\tilde f({\bf p})$ are square integrable functions in momentum space with respect 
to
the image of the relativistic invariant measure  $d{\bf p}/2\omega({\bf p})$, where $d{\bf p}$ 
is the Lebesgue
measure in momentum space. The scalar product on $F$ is defined by
$$
\left \langle f,g \right \rangle_F=(2 \pi)^3 \int \tilde f^{*}({\bf p}) \tilde g({\bf 
p})\frac{ d{\bf p}}{2 \omega ({\bf
p})}.$$
where we have normalized the Fourier transform in such a way
that
$$f({\bf x})=\int \exp{ (i {\bf p}.{\bf x} )}\tilde f({\bf p}) d{\bf p},$$
$$\tilde f({\bf p})=
(2 \pi)^{-3}\int \exp{ (-i {\bf p}.{\bf x}) }\tilde f({\bf x}) d{\bf x},$$
$$\int \exp{ (i {\bf p}.{\bf x} )}
d{\bf p}= (2\pi)^{3}\delta({\bf x}).$$
The scalar product on $F$ can be expressed also in the form
$$\left \langle f,g \right \rangle_F=\int\int f({\bf x}^\prime)^*W({\bf x}^\prime-{\bf 
x})g({\bf x})\ d{\bf x}^\prime d{\bf x},$$
where we have introduced the two point Wightman function at fixed time, defined by
$$W({\bf x}^\prime-{\bf x})=(2\pi)^{-3}\int \exp{ (i {\bf p}.({\bf x}^\prime-{\bf x})) }\frac{ 
d{\bf p}}{2 \omega ({\bf
p})}.$$ 
A unitary irreducible representation of the
Poincar\'e group can be defined on $F$ in the obvious way. In particular, the generators of
space translations are given by multiplication by the components of ${\bf p}$ in momentum 
space, and
the generator of time translations (the energy of the particle) is given by $\omega ({\bf 
p})$.

For the scalar product of time evolved wave functions we can write
$$\left \langle \exp(-it^\prime)f,\exp(-it)g \right \rangle_F=\int\int f({\bf 
x}^\prime)^*W(t^\prime-t,{\bf x}^\prime-{\bf x})g({\bf x})\ d{\bf x}^\prime d{\bf x},$$
where we have introduced the two point Wightman function, defined by
$$W(t^\prime-t,{\bf x}^\prime-{\bf x})=(2\pi)^{-3}\int \exp(-i(t-t^\prime))\exp{ (i {\bf 
p}.({\bf x}^\prime-{\bf x})) }\frac{ d{\bf p}}{2 \omega ({\bf
p})}.$$ 

To the
physical single particle system living in Minkowski space-time we associate a kind of 
mathematical
image, living in Euclidean space, from which all properties of the physical system can be 
easily
derived. We start from the two point Schwinger function
$$
S(x)=
\frac{1}{(2\pi)^4}\int \frac{\exp{ (i p \cdot x )}}{\sqrt{ p^2 + m^2}}\ dp,
$$
which is the
analytic continuation of the previously given two point Wightman function into the Schwinger 
points.
Here $x,p \in R^4$, and $p \cdot x=\sum_{i=1}^4 x_i p_i $. While $dp$ and $dx$ are the 
Lebesgue measures
in the $R^4$ momentum and configuration spaces respectively. The function $S(x)$ is positive 
and
analytic for $x\ne 0$, decreases as $\exp{(-m \parallel x\parallel)}$ as $x\to\infty$, and 
satisfies the
equation
$$(-\Delta + m^2 ) S(x)= \delta(x),$$
where $\Delta =
\sum_{i=1}^4 \frac{\partial^2}{\partial x_i^2}$ is the Laplacian in four dimensions.

The mathematical
image we are looking for is described by the Hilbert space $N$ of those tempered distributions 
in
four-dimensional configuration space $R^4$, whose Fourier transforms are square integrable 
with
respect to the measure $dp/\sqrt{ p^2 + m^2}$. The scalar product on $N$ is defined
by
$$\left\langle f,g \right\rangle_N=(2\pi)^4\int \tilde f^{*}(p) \tilde g(p)\frac{
dp}{\sqrt{ p^2 + m^2}}.$$
Four dimensional Fourier transform are normalized as
follows
$$
f(x) = \int \exp{( i p.x) }{\tilde f}(p)\ dp,$$
$${\tilde f}(p)  =  (2\pi)^{-4}\int \exp{( -i p.x) }\tilde f(x) dx,$$
$$\int \exp{( i p.x) } dp = (2\pi)^{4}\delta( x).
$$
We write also
$$\left\langle f,g \right\rangle_N=\int\int f^{*}(x)S(x-y)g(y) \ dx \ dy=
\left\langle f, (-\Delta + m^2 )^{-1} g \right\rangle,$$
where $\left \langle ,\right \rangle$ is the ordinary Lebesgue product defined on Fourier 
transforms and $(-\Delta + m^2
)^{-1}$ is given by multiplication by $(p^2 + m^2 )^{-1}$ in momentum space. The Schwinger 
function
$S(x-y)$ is formally the kernel of the operator $(-\Delta + m^2)^{-1}$. The Hilbert space $N$ 
is the
carrier space of a unitary (nonirreducible) representation of the four-dimensional Euclidean 
group
$E(4)$. In fact, let$(a,R)$ be an element of $E(4)$
\begin{eqnarray}
\nonumber (a,R): R^4 & \to & R^4\\
 \nonumber x & \to &
Rx+a,\end{eqnarray}where $a \in R^4$, and R is an orthogonal matrix, $R R^T=R^T R = 1_4$. Then 
the
transformation $u(a,R)$ defined by
\begin{eqnarray}\nonumber u(a,R) :  N & \to & N\\ \nonumber
f(x) & \to & (u(a,R)f)(x)=f(R^{-1}(x-a)),\end{eqnarray}
provides the representation. In particular, we
consider the reflection $r_0$ with respect to the hyperplane $x_4=0$, and the translations 
$u(t)$ in
the $x_4$ direction. Then we have $r_0 u(t) r_0 = u(-t)$, and analogously for other 
hyperplanes. 

Now
we introduce a local structure on $N$ by considering, for any closed region $A$ of $R^4$, the
subspace $N_A$ of $N$ made by distributions in $N$ with support on $A$. We call $e_A$ the 
orthogonal
projection on $N_A$. It is obvious that if $A\in B$ then $N_A\in N_B$ and $e_A e_B = e_B e_A 
=e_A$. A
kind of (pre)Markov property for one particle systems is introduced as follows. Consider a 
closed three
dimensional piece-wise smooth manifold $\sigma$, which divides $R^{4}$ in two closed regions 
$A$ and
$B$, having $\sigma$ in common. Therefore $\sigma\in A$, $\sigma\in B$, $A\cap B =\sigma$,
$A \cup B =R^4$.
Let $N_A$, $N_B$, $N_\sigma$, and $e_A$, $e_B$, $e_\sigma$ be the associated subspaces and
projections respectively. Then $N_\sigma \subset N_A$,$N_\sigma \subset N_B$, and $e_\sigma 
e_A = e_A e_\sigma =
e_\sigma$, $e_\sigma e_B= e_B e_\sigma = e_\sigma$. It is very simple to prove the
following.\begin{theorem}\label{preMarkov}Let $e_A$, $e_B$, $e_\sigma$ be defined as above, 
then
$e_A e_B = e_B e_A =e_\sigma$.\end{theorem}Clearly, it is enough to show that for any $f\in N$ 
we
have $ e_A e_B f \in N_\sigma$. In that case $ e_\sigma e_A e_B f= e_A e_B f$, from which the 
theorem
easily follows. Since $ e_A e_B f$ has support on $A$, we must show that for any 
$C^{\infty}_0$
function g with support on $A_\sigma$ we have $\left\langle g, e_A e_B f \right\rangle=0$. 
Then $ e_A
e_B f$ has support on $\sigma$, and the proof is complete. Now we have
\begin{eqnarray}\nonumber \left\langle g,
e_A e_B f \right\rangle =& \left\langle (-\Delta + m^2)g,e_A e_B f \right\rangle_N\\ 
\nonumber=& \left\langle
e_A (-\Delta + m^2)g, e_B f\right\rangle_N\\ \nonumber=& \left\langle (-\Delta + m^2)g, e_B f
\right\rangle_N\\ \nonumber =& \left\langle g, e_B f \right\rangle = 0,\end{eqnarray}
where we have used the
definition of $\left\langle \right\rangle_N$ interms of $\left\langle \right\rangle$, the fact 
that
$ e_A (-\Delta +m^2)g=(-\Delta + m^2)g$, since $(-\Delta + m^2)g$ has support on $A_\sigma$, 
and the
fact that $ e_B f $ has support on $B$. This ends the proof of the (pre))Markov property for 
one particle
systems. 

A very important role in the theory is played by subspaces of $N$ associated to hyperplanes
in $R^4$. To fix ideas, consider the hyperplane $x_4=0$, and the associated subspace $N_0$. A
tempered distribution in $N$ with support on $x_4=0$ has necessarily the form
$(f \otimes\delta_0)(x) \equiv f({\bf x})\delta(x_4)$, with $f \in F$. By using the basic 
magic formula, for $x \ge 0$ and
$M>0$, $$\int_{-\infty}^{+\infty}\frac{\exp (i p x )}{p^2 + M^2} dp = \frac{\pi}{M} 
\exp{(-Mx)},$$ it is
immediate to verify that $ \parallel f \otimes \delta_0\parallel_N = \parallel f \parallel_F$. 
Therefore, we have an isomorphic and isometric
identification of the two Hilbert spaces $F$ and $N_0$. Obviously, similar considerations hold 
for
any hyperplane. In particular, we consider the hyperplanes $x_4=t$, and the associated 
subspaces
$N_t$. Let us introduce injection operators $j_t$ defined by
\begin{eqnarray}\nonumber j_t: F & \to N\\ \nonumber f & \to f\otimes
\delta_t,\end{eqnarray} where $f$ is a generic element of $F$, with values $f({\bf x})$, and
$(f\otimes\delta_t)(x)= f({\bf x})\delta(x_4-t)$. It is immediate to verify the following 
properties for $j_t$
and its adjoint $j_t^{*}$. The range of $j_t$ is $N_t$. Moreover, $j_t$ is an isometry, so 
that
$j_t^{*} j_t=1_F$, $j_tj_t^{*}=e_t$, where $1_F$ is the identity on $F$, and $e_t$ is the 
projection
on $N_t$. Moreover, $ e_t j_t = j_t$ and $ j_t^{*}= j_t^{*}e_t$.

If we introduce translations $u(t)$
along the $x_4$ direction and the reflection $r_0$ with respect to $x_4=0$, then we also the
covariance property $u(t) j_s=j_{t+s}$, and the reflexivity property $r_0 j_0 =j_0$, 
$j_0^{*}r_0=j_0^{*}$. The reflexivity property is very important. It tells us that $r_0$ 
leaves $N_0$ pointwise invariant, and it is an immediate consequence of the fact that 
$\delta(x_4)=\delta(-x_4)$.

Therefore, if we start from $N$ we can obtain $F$, by taking the projection $j_\pi$ with 
respect to some hyperplane $\pi$, in particular $x_4=0$. It is also obvious that we can induce 
on $F$ a representation of $E(3)$ by taking those element of $E(4)$ that leave $\pi$ 
invariant. 

Let us now see how we can define the Hamiltonian on $F$ starting from properties of $N$. Since 
we are considering the simple case of the one particle system, we could just perform the 
following construction explicitly by hands, through a simple application of the basic magic 
formula given before. But we prefer to follow a route that emphasizes Markov property and can 
be immediately generalized to more complicated cases.

Let us introduce the operator $p(t)$ on $F$ defined by the dilation $p(t)=j_0^* j_t=j_0^* u(t) 
j_0$, $t\ge 0$. Then we prove the following.
\begin{theorem} \label{semigroup}
The operator $p(t)$ is bounded and selfadjoint. The family $\{p(t)\}$, for $t \ge 0$, is a 
norm-continuous semigroup.
\end{theorem}
Proof. Boundedness and continuity are obvious. Selfadjointness is a consequence of 
reflexivity. In fact $$p^*(t)=j_0^* u(-t)j_0=j_0^* r_0 u(t) r_0 j_0 = j_0^* u(t) j_0 = p(t).$$
The semigroup property is a consequence of the Markov property. In fact, let us introduce 
$N_+$, $N_0$, $N_-$ as subspaces of $N$ made by distributions with support in the regions 
$x_4\ge 0$, $x_4=0$, $x_4 \le 0$, respectively, and call $e_+$, $e_0$, $e_-$ the respective 
projections. By Markov property we have $e_0=e_- e_+$. Now write for $s,t\ge 0$
$$p(t)p(s)=j_0^* u(t) j_0 j_0^* u(s) j_0 = j_0^* u(t) e_0 u(s) j_0.$$
If $e_0$ could be cancelled, then the semigroup property would follow from the group property 
of the translations $u(t) u(s)= u(t+s)$ (a miracle of the dilations!). For this, consider the 
matrix element 
$$\left \langle f,p(t)p(s)g \right \rangle_F=\left \langle u(-t)j_0 f, e_0 u(s)j_0 g \right 
\rangle_N,$$
recall $e_0=e_- e_+$, and use $u(s)j_0 g \in N_+$, and $u(-t)j_0 f \in N_-$.

Let us call $h$ the generator of $p(t)$, so that $p(t)= \exp (-th)$, for $t \ge 0$. By 
definition, $h$ is the Hamiltonian of the physical system. A simple explicit calculation shows 
that $h$ is just the energy $ \omega$ introduced before. Starting from the representation of 
the Euclidean group $E(3)$ already given and from the Hamiltonian, we immediately get a 
representation of the full Poincar\'e group on $F$. Therefore, all physical properties of the 
one particle system have been reconstructed from its Euclidean image on the Hilbert space $N$.

As a last remark of this section, let us note that we can consider the real Hilbert spaces 
$N_r$ and $F_r$, made of real elements (in configuration space) in $N$ and $F$. The operators 
$u(a,t)$, $u(t)$, $r_0$, $j_\pi$, $j^*_\pi$, $e_A$ are all reality preserving, \textit{i.e.} 
they map real spaces into real spaces.

This exhausts our discussion about the one particle system. For more details we refer to 
\cite{GRS} and \cite{Simon}. We have introduced the Euclidean image, discussed its main 
properties, and shown how we can derive all properties of the physical system from its 
Euclidean image. In the next sections, we will show how this kind of construction carries 
through the second quantized case and the interacting case.

\section{Second quantization and free fields}

We begin this section with a short review about the procedure of second quantization based on 
probabilistic methods, by following mainly Nelson \cite{Nel1}, see also \cite{GRS} and 
\cite{Simon}. Probabilistic methods are particularly useful in the frame of the Euclidean 
theory. 

Let ${\cal H}$ be a real Hilbert space, with symmetric scalar product $\left \langle \ ,\right 
\rangle$. Let $\phi(u)$ the elements of a family of centered Gaussian random variables indexed 
by $u \in {\cal H}$, uniquely defined by the expectation values
$E(\phi(u))=0$, $E(\phi(u)\phi(v))=\left \langle u,v \right \rangle$. Since $\phi$ is Gaussian 
we also have
$$E(\exp(\lambda \phi(u)))=\exp(\frac{1}{2}\lambda^2 \left \langle u,u \right \rangle),$$ and
$$E(\phi(u_1)\phi(u_2) \dots \phi(u_n) )= [ u_1 u_2 \dots u_n ].$$
Here $[\dots ]$ is the Hafnian of elements $[u_i u_j ]= \left \langle u_i,u_j \right \rangle$, 
defined to be zero for odd $n$, and for even $n$ by the recursive formula
$$[ u_1 u_2 \dots u_n ]=\sum_{i=2}^n [u_1 u_i][ u_1 u_2 \dots u_n ]^{\prime},$$
where in $[\dots ]^{\prime}$ the terms $u_1$ and $u_i$ are suppressed. Hafnians, from the 
Latin name of Copenhagen, the first seat of the theoretical group of CERN, were introduced in 
quantum field theory by Caianiello \cite{Caia}, as a useful tool when dealing with Bose 
statistics.

Let $(Q,\Sigma,\mu)$ be the underlying probability space where $\phi$ are defined as random 
variables. Here $Q$ is a compact space, $\Sigma$ a $\sigma$-algebra of subsets of $Q$, and 
$\mu$ a regular, countable additive probability measure on $\Sigma$, normalized to 
$\mu(Q)=\int_Q d \mu =1$. 

The fields $\phi(u)$ are represented by measurable functions on $Q$. The probability space is 
uniquely defined, but for trivial isomorphisms, if we assume that $\Sigma$ is the smallest 
$\sigma$-algebra with respect to which all fields $\phi(u)$, with $u \in {\cal H}$, are 
measurable. Since $\phi(u)$ are Gaussian, then they are represented by $L^p (Q,\Sigma,\mu)$ 
functions, for any $p$ with $1 \le p < \infty$, and the expectations will be given by
$$E(\phi(u_1)\phi(u_2) \dots \phi(u_n) )=\int_Q \phi(u_1)\phi(u_2) \dots \phi(u_n) \ d\mu,$$ 
where, by a mild abuse of notation, the $\phi(u_i)$ in the right hand side denote the $Q$ 
space functions which represent the random variables $\phi(u_i)$. We call the complex Hilbert 
space ${\cal F}=\Gamma ({\cal H})=L^2 (Q,\Sigma,\mu)$ the $\Phi o \kappa$ space constructed on 
${\cal H}$, and the function $\Omega_0 \equiv 1$ on $Q$ the $\Phi o \kappa$ vacuum.

In order to introduce the concept of second quantization of operators, we must introduce 
subspaces of ${\cal F}$ with a ``fixed number of particles''. Call ${\cal F}_{(0)}=\{ \lambda 
\Omega_0 \}$, where $\lambda$ is any complex number. Define ${\cal F}_{(\le n)}$ as the 
subspace of ${\cal F}$ generated by complex linear combinations of monomials of the type 
$\phi(u_1) \dots \phi(u_j)$, with $u_i \in {\cal H}$, and $j \le n$. Then ${\cal F}_{(\le 
n-1)}$ is a subspace of ${\cal F}_{(\le n)}$. We define ${\cal F}_{(n)}$, the $n$ particle 
subspace, as the orthogonal complement of ${\cal F}_{(\le n-1)}$ in ${\cal F}_{(\le n)}$, so 
that
$${\cal F}_{(\le n)}={\cal F}_{(n)}\oplus{\cal F}_{(\le n-1)}.$$
By construction the ${\cal F}_{(n)}$ are orthogonal, and it is not difficult to verify that 
$${\cal F}=\oplus_{n=0}^\infty {\cal F}_{(n)}.$$

Let us now introduce Wick normal products by the definition
$$:\phi(u_1)\phi(u_2) \dots \phi(u_n):=E_{(n)}\phi(u_1)\phi(u_2) \dots \phi(u_n),$$       
where $E_{(n)}$ is the projection on ${\cal F}_{(n)}$. It is not difficult to prove the usual 
Wick theorem (see for example \cite{GRS}), and its inversion given by Caianiello \cite{Caia}.

It is interesting to remark that, in the frame of the second quantization performed with 
probabilistic methods, it is not necessary to introduce creation and destruction operators as 
in the usual treatment. However the two procedures are completely equivalent, as shown for 
example in \cite{Simon}.

Given an operator $A$ from the real Hilbert space ${\cal H}_1$ to the real Hilbert space 
${\cal H}_2$, we define its second quantized operator $\Gamma(A)$ through the following 
definitions
$$\Gamma(A) \Omega_{01}=\Omega_{02},$$  
$$\Gamma(A):\phi_1 (u_1)\phi_1 (u_2) \dots \phi_1 (u_n):=:\phi_2 (A u_1)\phi_2 (Au_2) \dots 
\phi_2 (Au_n):,$$
where we have introduced the probability spaces $Q_1$ and $Q_2$, their vacua $\Omega_{01}$ and 
$\Omega_{02}$, and the random variables $\phi_1$ and $\phi_2$, associated to ${\cal H}_1$ and 
${\cal H}_2$, respectively. The following remarkable theorem by Nelson \cite{Nel1} gives a 
full characterization of $\Gamma(A)$, very useful in the applications.
\begin{theorem}\label{hyper}   
Let $A$ be a contraction from the real Hilbert space ${\cal H}_1$ to the real Hilbert space 
${\cal H}_2$. Then $\Gamma(A)$ is an operator from $L^1 _{(1)}$ to $L^1 _{(2)}$ which is 
positivity preserving, $\Gamma(A)u\ge =0$ if $u\ge =0$, and such that $E(\Gamma(A)u)=E(u)$. 
Moreover, $\Gamma(A)$ is a contraction from $L^p_{(1)}$ to $L^p_{(2)}$ for any $p$, $1\le p < 
\infty$. Finally, $\Gamma(A)$ is also a contraction from $L^p_{(1)}$ to $L^q_{(2)}$, with $q 
\ge p$, if $\parallel A \parallel^2 \le (p-1)/(q-1)$.
\end{theorem}

We have indicated with $L^p_{(1)}$, $L^p_{(2)}$ the $L^p$ spaces associated to ${\cal H}_1$ 
and ${\cal H}_2$, respectively. This is the celebrated Nelson best hypercontractive estimate. 
For the proof we refer to the original paper \cite{Nel1}, see also \cite{Simon}.

This exhausts our short review on the theory of second quantization based on probabilistic 
methods.

The usual time-zero quantum field $\bar\phi (u)$, $u\in F_r$, in the $\Phi o \kappa$ 
representation, can be obtained through second quantization starting from $F_r$. We call 
$(\bar Q, \bar\Sigma, \bar\mu)$ the underlying probability space, and ${\cal F}=\Gamma 
(F_r)=L^2 (\bar Q,\bar\Sigma,\bar\mu)$ the Hilbert $\Phi o \kappa$ space of the free physical 
particles.

Now we introduce the free Markov field $\phi(f)$, $f\in N_r$, by taking $N_r$ as the starting 
point. We call $(Q,\Sigma,\mu)$ the associated probability space. We introduce the Hilbert 
space ${\cal N}=\Gamma (N_r)=L^2 (Q,\Sigma,\mu)$, and the operators $U(a,R)=\Gamma (u(a,R))$, 
$R_0=\Gamma (r_0)$, $U(t)=\Gamma (u(t))$, $E_A=\Gamma (e_A)$, and so on, for which the 
previous Nelson theorem holds (take ${\cal H}_1={\cal H}_2=N_r$). 

Since in general $\Gamma(AB)=\Gamma(A)\Gamma(B)$, then we have immediately the following 
expression of the Markov property $E_\sigma=E_A E_B$, where the closed regions $A$, $B$, 
$\sigma$ of the Euclidean space have the same properties explained before during the proof of 
the (pre)Markov property for one particle systems.

It is obvious that $E_A$ can also be understood as conditional expectation with respect to the 
sub-$\sigma$-algebra $\Sigma_A$ generated by the field $\phi(f)$ with $f\in N_r$  and the 
support of $f$ on $A$.

The relation, previously pointed out, between $N_t$ subspaces and $F$ are also valid for their 
real parts $N_{rt}$ and $F_r$. Therefore they carry out through the second quantization 
procedure. We introduce $J_t=\Gamma(j_t)$ and $J_t^*=\Gamma(j_t^*)$, then the following 
properties hold. $J_t$ is an isometric injection of $L^p (\bar Q,\bar\Sigma,\bar\mu)$ into 
$L^p (Q,\Sigma,\mu)$, the range of $J_t$ as an operator $L^2\to L^2$ is obviously ${\cal 
N}_t=\Gamma(N_{rt})$, moreover $J_t J_t^*=E_t$. The free Hamiltonian $H_0$ is given for $t\ge 
0$ by 
$$J_0^* J_t =\exp(-t H_0)=\Gamma(\exp(-t\omega)).$$
Moreover, we have the covariance property $U(t)J_0=J_t$, and the reflexivity $R_0 J_0=J_0$, 
$J_0^* R_0 = J_0^*$.

These relations allow a very simple expression for the matrix elements of the Hamiltonian 
semigroup in terms of Markov quantities. In fact, for $u,v\in {\cal F}$ we have
$$\left \langle u,\exp(-t H_0) v\right \rangle=\int_Q (J_t u)^* J_0 v \ d\mu.$$

In the next section we will generalize this representation to the interacting case.

Finally, let us derive the hypercontractive property of the free Hamiltonian semigroup.

Since $\parallel \exp(-t\omega)\parallel \ \le\exp(-tm)$, where $m$ is the mass of the 
particle, we have immediately by a simple application of Nelson theorem
$$\parallel\exp(-tH_0)\parallel_{p,q} \ \le 1,$$                  
provided $q-1\le(p-1)\exp(2tm)$, where $\parallel \dots \parallel_{p,q}$ denotes the norm of 
an operator from $L^p$ to $L^q$ spaces.

\section{Interacting fields}

The discussion of the previous sections was limited to free fields both in Minkowski and 
Euclidean spaces. Now we must introduce interaction in order to get nontrivial theories.

Firstly, as a general motivation, we will proceed quite formally, then we will resort to 
precise statements.

Let us recall that in standard quantum field theory, for scalar self-coupled fields, the time 
ordered products of quantum fields in Minkowski space-time can be expressed formally through 
the formula
$$\frac{\left \langle T(\phi(x_1)\dots \phi(x_n)\exp{(i\int {\cal L}\ dx)})\right \rangle}
{\left \langle T\exp{(i\int {\cal L}\ dx)}\right \rangle},$$
where $T$ denotes time ordering, $\phi$ are free fields in Minkowski space-time, ${\cal L}$ is 
the interaction Lagrangian, and $\left \langle\dots\right \rangle$ are vacuum averages. As it 
is very well known, this expression can be put for example at the basis of perturbative 
expansions, giving rise to terms expressed through Feynman graphs. The appropriate chosen 
normalization provides automatic cancellation of the vacuum to vacuum graphs.

Now we can introduce a formal analytic continuation to the Schwinger points, as previously 
done for the one particle system, and obtain the following expression for the analytic 
continuation of the field time ordered products, now called Schwinger functions,
$$S(x_1,\dots,x_n)=\frac{\left \langle \phi(x_1)\dots \phi(x_n)\exp U \right \rangle}
{\left \langle \exp U \right \rangle}.$$
Here $x_1,\dots,x_n$ denote points in Euclidean space, $\phi$ are the Euclidean fields 
introduced before. The chronological time ordering disappears, because the fields $\phi$ are 
commutative, and there is no distinguished ``time'' direction in Euclidean space. The symbol 
$\left \langle\dots\right \rangle$ denotes here expectation values represented by $\int\dots\ 
d\mu$, as explained before, and $U$ is the Euclidean ``action'' of the system formally given 
by the integral on Euclidean space
$$U=-\int P(\phi(x))\ dx,$$     
if the field self-interaction is produced by the polynomial $P$.

Therefore, these formal considerations suggest that the passage from the free Euclidean theory 
to the fully interacting one is obtained through a change of the free probability measure 
$d\mu$ to the interacting measure
$$\exp U\ d\mu / \int_Q \exp U\ d\mu.$$
The analogy with classical statistical mechanics is evident. The expression $\exp U$ acts as 
\textit{Boltzmannfaktor}, and $Z=\int_Q \exp U\ d\mu$ is the partition function.

Our task will be to make these statements precise from a mathematical point of view. We will 
be obliged to introduce cutoffs, and then be involved in their careful removal.
 
For the sake of convenience, we make the substantial simplification of considering only 
two-dimensional theories (one space - one time dimensions in the Minkowski region) for which 
the well known ultraviolet problem of quantum field theory gives no trouble. There is no 
difficulty in translating the content of the previous sections to the two-dimensional case.

Let $P$ be a real polynomial, bounded below and normalized to $P(0)=0$. We introduce 
approximations $h$ to the Dirac $\delta$ function at the origin of the two-dimensional 
Euclidean space $R^2$, with $h\in N_r$. Let $h_x$ be the translate of $h$ by $x$, with $x\in 
R^2$. The introduction of $h$, equivalent to some ultraviolet cutoff, is necessary, because 
local fields, of the formal type $\phi(x)$, have no rigorous meaning, and some smearing is 
necessary.

For some compact region $\Lambda$ in $R^2$, acting as space cutoff (infrared cutoff), 
introduce the $Q$ space function
$$U^{(h)}_{\Lambda}=-\int_\Lambda :P(\phi(h_x)):\ dx,$$  
where $dx$ is the Lebesgue measure in $R^2$. It is immediate to verify that 
$U^{(h)}_{\Lambda}$ is well defined, bounded below and belongs to $L^p(Q,\Sigma,\mu)$, for any 
$p$, $1\le p<\infty$. This is the infrared and ultraviolet cutoff action. Notice the presence 
of the Wick normal products in its definition. They provide a kind of automatic introduction 
of counter-terms, in the frame of renormalization theory.

The following theorem allows to remove the ultraviolet cutoff.
\begin{theorem}\label{U}
Let $h\to\delta$, in the sense that the Fourier transforms $\tilde h$ are uniformly bounded 
and converge pointwise in momentum space to the Fourier transform of the $\delta$ function 
given by $(2\pi)^{-2}$. Then $U^{(h)}_{\Lambda}$ is $L^p$ convergent for any $p$, $1\le 
p<\infty$, as $h\to\delta$. Call $U_{\Lambda}$ the $L^p$ limit, then $U_{\Lambda},\exp 
U_{\Lambda}\in L^p(Q,\Sigma_\Lambda,\mu)$, for $1\le p<\infty$.
\end{theorem}

The proof uses standard methods of probability theory, and originates from pioneering work of 
Nelson in \cite{Nel3}. It can be found for example in \cite{GRS}, and \cite{Simon}.

Since $U_{\Lambda}$ is defined with normal products, and the interaction polynomial $P$ is 
normalized to $P(0)=0$, an elementary application of Jensen inequality gives
$$\int_Q \exp U_\Lambda\ d\mu \ge \exp \int_Q U_\Lambda\ d\mu =1.$$

Therefore, we can rigorously define the new space cutoff measure in $Q$ space
$$d\mu_\Lambda=\exp U_\Lambda\ d\mu / \int_Q \exp U_\Lambda\ d\mu.$$

The space cutoff interacting Euclidean theory is defined by the same fields on $Q$ space, but 
with a change in the measure and therefore in the expectation values. The correlations for the 
interacting fields $\bar\phi$ are the cutoff Schwinger functions
$$S_\Lambda(x_1,\dots,x_n)=\left\langle \bar\phi(x_1)\dots\bar\phi(x_n)  
\right\rangle=Z_\Lambda^{-1}\left\langle \phi(x_1)\dots\phi(x_n)  \exp 
U_\Lambda\right\rangle,$$
where the partition function is
$$Z_\Lambda=\left\langle \exp U_\Lambda\right\rangle.$$

We see that the analogy with statistical mechanics is complete here. Of course, the 
introduction of the space cutoff $\Lambda$ destroys translation invariance. The full Euclidean 
covariant theory must be recovered by taking the infinite volume limit $\Lambda\to R^2$ on 
field correlations. For the removal of the space cutoff all methods of statistical mechanics 
are available. In particular, correlation inequalities of ferromagnetic type can be easily 
exploited, as shown for example in \cite{GRS} and \cite{Simon}.

We would like to conclude this section by giving the connection between the space cutoff 
Euclidean theory and the space cutoff Hamiltonian theory in the physical $\Phi o \kappa$ 
space. 

For $\ell\ge 0$, $t\ge 0$, consider the rectangle in $R^2$      
$$\Lambda(\ell,t)=\{(x_1,x_2): -\frac{\ell}{2}\le x_1 \le \frac{\ell}{2}, 0\le x_2 \le t \},$$
and define the operator in the physical $\Phi o \kappa$ space
$$P_{\ell}(t)=J_0^* \exp U_\Lambda(\ell,t) J_t,$$ 
where $J_0$ and $J_t$ are injections relative to the lines $x_2=0$ and $x_2=t$, respectively. 
Then the following theorem, largely due to Nelson, holds.
\begin{theorem}\label{FKN}
The operator $P_{\ell}(t)$ is bounded and selfadjoint. The family $\{P_{\ell}(t)\}$, for 
$\ell$ fixed and $t\ge 0$ is a strongly continuous semigroup. Let $H_\ell$ be its lower 
bounded selfadjoint generator, so that $P_{\ell}(t)=\exp(-t H_\ell)$. On the physical $\Phi o 
\kappa$ space, there is a core ${\cal D}$ for $H_\ell$ such that on ${\cal D}$ the following 
equality holds $H_\ell=H_0+V_\ell$, where $H_0$ is the free Hamiltonian introduced before and 
$V_\ell$ is the volume cutoff interaction  given by
$$V_\ell=\lim \int_{-\frac{\ell}{2}}^\frac{\ell}{2} :P(\bar\phi(h_{x_1})):\ dx_1,$$
where $h_{x_1}$ are the translates of approximations to the $\delta$ function at the origin on 
the $x_1$ space, and the limit is taken in $L^p$, in analogy to what has been explained for 
the two-dimensional case in the definition of $U_\Lambda$.
\end{theorem}     

While we refer to \cite{GRS} and \cite{Simon} for a full proof, we mention here that 
boundedness is related to hypercontractivity of the free Hamiltonian, selfadjointness is a 
consequence of reflexivity, and the semigroup property follows from Markov property. This 
theorem is remarkable, because it expresses the cutoff interacting Hamiltonian semigroup in an 
explicit form in the Euclidean theory through probabilistic expectations. In fact we have
$$\left \langle u,\exp(-t H_\ell) v\right \rangle=\int_Q (J_t u)^* J_0 v \exp 
U_\Lambda(\ell,t)\ d\mu.$$

We could call this expression as the Feynman-Kac-Nelson formula, in fact it is nothing but a 
path integral expressed in stochastic terms, and adapted to the Hamiltonian semigroup.

By comparison with the analogous formula given for the free Hamiltonian semigroup, we see that 
the introduction of the interaction inserts the \textit{Boltzmannfaktor} under the integral.

As an immediate consequence of the Feynman-Kac-Nelson formula, together with Euclidean 
covariance, we have the following astonishing Nelson symmetry
$$\left \langle \Omega_0,\exp(-t H_\ell) \Omega_0\right \rangle=\left \langle 
\Omega_0,\exp(-\ell H_t) \Omega_0\right \rangle,$$   
which was at the basis of \cite{G} and \cite{GRS1}, and played some role in showing the 
effectiveness of Euclidean methods in constructive quantum field theory.

It is easy to establish, through simple probabilistic reasoning, that $H_\ell$ has a unique 
ground state $\Omega_\ell$ of lowest energy $E_\ell$. For a convenient choice of normalization 
and phase factor, one has $\parallel \Omega_\ell \parallel_2=1$, and $\Omega_\ell > 0$ almost 
everywhere on $Q$ space (for Bosonic systems ground states have no nodes in configuration 
space!). Moreover, $\Omega_\ell \in L^p$, for any $1\le p <\infty$. If $\ell>0$ and the 
interaction is not trivial, then $\Omega_\ell \neq \Omega_0$, $E_\ell<0$, and $\parallel 
\Omega_\ell \parallel_1<1$. Obviously $\parallel\exp(-t H_\ell)\parallel_{2,2}=\exp(-t 
E_\ell)$.

The general structure of Euclidean field theory, as explained in this section, has been at the 
basis of all applications in constructive quantum field theory. These applications include the 
proof of the existence of the infinite volume limit, with the establishment of all Wightman 
axioms, for two and three-dimensional theories. Moreover, the existence of phase transitions 
and symmetry breaking has been firmly established. Extensions have been also given to theories 
involving Fermions, and to gauge field theory. Due to the scope of this review, limited to a 
description of the general structure of Euclidean field theory, we can not give a detailed 
treatment of these applications. Therefore, we refer to recent general reviews on constructive 
quantum field theory for a complete description of all results, as for example explained by 
Jaffe in \cite{Jaffe}. For recent applications of Euclidean field theory to quantum fields on 
curved space-time manifolds we refer for example to \cite{Schli}.

\section{The physical interpretation of Euclidean field theory}

Euclidean field theory has been considered by most researchers as a very useful tool for the 
study of quantum field theory. In particular, it is quite easy for example to obtain the fully 
interacting Schwinger functions in the infinite volume limit in two-dimensional space-time. At 
this point the problem arises of connecting these Schwinger functions with observable physical 
quantities in Minkowski space-time. A very deep result of Osterwalder and Schrader \cite{OS} 
gives a very natural interpretation of the resulting limiting theory. In fact, the Euclidean 
theory, as has been shown before, arises from an analytic continuation from the physical 
Minkowski space-time to the Schwinger points, through a kind of analytic continuation in time, 
also called Wick rotation, because Wick exploited this trick in the study of the 
Bethe-Salpeter equation. Therefore, having obtained the Schwinger functions for the full 
covariant theory, after all cutoff removal, it is very natural to try to reproduce the inverse 
analytic continuation in order to recover the Wightman functions in Minkowski space-time. 
Therefore, Osterwalder and Schrader have been able to identify a set of conditions, quite easy 
to verify, wich allow to recover Wightman functions from Schwinger functions. A key role in 
this reconstruction theorem is played by the so called reflection positivity for Schwinger 
functions, a property quite easy to verify. In this way a fully satisfactory solution for the 
physical interpretation of Euclidean field theory is achieved.

From an historical point of view, an alternate route is possible. In fact, at the beginning of 
the exploitation of Euclidean methods in constructive quantum field theory, Nelson was able to 
isolate a set of axioms for the Euclidean \textit{fields} \cite{Nel}, allowing the 
reconstruction of the physical theory. Of course, Nelson axioms are more difficult to verify, 
since they involve properties of the Euclidean fields and not only of the Schwinger functions. 
However, it is still very interesting to investigate whether the Euclidean fields play only an 
auxiliary role in the construction of the physical content of relativistic theories, or do 
they have a more fundamental meaning.

From a physical point of view, the following considerations could also lead to further 
developments along this line. By its very structure, the Euclidean theory contains the fixed 
time quantum correlations in the vacuum. In elementary quantum mechanics, it is possible to 
derive all physical content of the theory from the simple knowledge of the ground state wave 
function, including scattering data. Therefore, at least in principle, it should be possible 
to derive all physical content of the theory directly from the Euclidean theory, without any 
analytic continuation.

We conclude this short section on the physical interpretation of the Euclidean theory, with a 
mention to a quite surprising result \cite{GR}, obtained by submitting classical field theory 
to the procedure of stochastic quantization in the sense of Nelson \cite{Nelsonbook}. The 
procedure of stochastic quantization associates a stochastic process to each quantum state. In 
this case, in a fixed reference frame, the procedure of stochastic quantization applied to 
interacting fields, produces for the ground state a process, in the physical space-time, which 
has the same correlations as Euclidean field theory. This open the way to a possible 
interpretation of Euclidean field theory directly in Minkowski space-time. However, a 
consistent development along this line requires a new formulation of representations of the 
Poincar\'e group in the form of measure preserving transformations in the probability space 
where the Euclidean fields are defined. This difficult task has not been accomplished yet.    

Research connected with this work was supported in part by MIUR (Italian Minister of 
Instruction, University and
Research), and by INFN (Italian National Institute for Nuclear
Physics).


\begin{thebibliography}{0}

\bibitem{Schwinger} J. Schwinger, 
\textit{On the Euclidean structure of relativistic field theory},
Proc. Nat. Acad. Sc. {\bf 44}, 956 (1958).

\bibitem{Nakano} T. Nakano,
\textit{Quantum field theory in terms of Euclidean parameters}, 
Prog. Theor. Phys. {\bf 21}, 241 (1959).

\bibitem{SW} R. Streater and A.S. Wightman, 
\textit{PCT, Spin and Statistics and All That}, Benjamin Scientific Publishers, New York, 
1964.

\bibitem{Sym} K. Symanzik, \textit{Euclidean quantum field theory, I, Equations for a scalar 
model},
J. Math. Phys. {\bf 7}, 510 (1966).

\bibitem{Sym1} K. Symanzik, \textit{Euclidean quantum field theory},
in ``Local Quantum Theory'' (ed. R. Jost), Academic Press, New York, 1969.

\bibitem{GRS} F. Guerra, L. Rosen and B. Simon, \textit{The $P(\phi)_2$ Euclidean quantum 
field theory as classical statistical mechanics}, Ann. Math. {\bf 101}, 111 (1975).

\bibitem{Wigh} A.S. Wightman, 
\textit{An introduction to some aspects of the relativistic dynamics of quantized fields},
in ``1964 Carg\'ese Summer School Lectures'' (ed. M. Levy), Gordon and Breach, New York, 1967.  

\bibitem{GJ} J. Glimm and A. Jaffe, \textit{Quantum Physics, A Functional Integral Point of 
View}, Springer Verlag, Berlin, 1981.

\bibitem{Nel} E. Nelson,
\textit{Construction of quantum fields from Markoff fields}, J. Funct. Anal. {\bf 12}, 97 
(1973).

\bibitem{Nel1} E. Nelson, 
\textit{The free Markoff field}, J. Funct. Anal. {\bf 12}, 211 (1973).

\bibitem{G} F. Guerra, 
\textit{Uniqueness of the vacuum energy density and van Hove phenomenon in the infinite volume 
limit for two-dimensional self-coupled Bose fields}, 
Phys. Rev. Lett. {\bf 28}, 1213 (1972).

\bibitem{GRS1} F. Guerra, L. Rosen and B. Simon, \textit{Nelson's symmetry and the infinite 
volume behavior of the vacuum in $P(\phi)_2$}, Commun. Math. Phys. {\bf 27}, 10 (1972).

\bibitem{OS} K. Osterwalder and R. Schrader, 
\textit{Axioms for Euclidean Green's functions},
Commun. Math. Phys. {\bf 31}, 83 (1973).

\bibitem{GR} F. Guerra and P. Ruggiero, \textit{New interpretation of the Euclidean Markov 
field in the framework of physical Minkowski space-time}, Phys. Rev. Lett. {\bf 31}, 1022 
(1973).

\bibitem{Simon}  B. Simon, 
\textit{The $P(\phi)_2$ Euclidean (Quantum) Field Theory},
Princeton University Press, Princeton, New Jersey, 1974.
 
 \bibitem{Caia}  E.R. Caianiello, 
\textit{Combinatorics and Renormalization in Quantum Field Theory},
 W.A. Benjamin, Inc., Reading, Massachusetts, 1973.
 
 \bibitem{Nel3}  E. Nelson, 
\textit{A quartic interaction in two dimensions},
 in ``Conference on the Mathematical Theory of Elementary Particles'', R. Goodman and I. 
Segal, Eds., MIT Press, Cambridge, Massachusetts, 1966. 
 
 \bibitem{Jaffe}  A. Jaffe, 
\textit{Constructive quantum field theory},
available on http://www.arthurjaffe.com/ .
 
 \bibitem{Schli}  D. Schlingemann, 
\textit{Euclidean field theory on a sphere},
 available on http://arXiv.org/abs/hep-th/9912235 .
 
 \bibitem{Nelsonbook}  E. Nelson, 
\textit{Quantum fluctuations},
 Princeton University Press, Princeton, New Jersey, 1985.
 
 
 
 
 
 


\end{thebibliography}
\end{document}